\documentclass[fleqn,usenatbib]{mnras}
\usepackage{newtxtext,newtxmath}

\usepackage[T1]{fontenc}
\usepackage{upgreek}
\usepackage{setspace}

\DeclareRobustCommand{\VAN}[3]{#2}
\let\VANthebibliography\thebibliography
\def\thebibliography{\DeclareRobustCommand{\VAN}[3]{##3}\VANthebibliography}

\usepackage{soul}

\usepackage{graphicx}	
\usepackage{amsmath}	

\usepackage{amssymb}	
\usepackage{breakurl}

\def\degr{\hbox{$^\circ$}}
\def\arcmin{\hbox{$^\prime$}}

\def\fdg{\hbox{$.\!\!^\circ$}}
\def\farcm{\hbox{$.\mkern-4mu^\prime$}}
\DeclareRobustCommand{\ion}[2]{\textup{#1\,\textsc{\lowercase{#2}}}}

\title[Revisiting G213.0$-$0.6]{G213.0$-$0.6, a true supernova remnant
  or just an \ion{H}{II} region?}

\author[X. Y. Gao et al.]{
X. Y. Gao,$^{1,4,5}$\thanks{E-mail: xygao@nao.cas.cn (XYG)}
C. J. Wu,$^{1,4,5}$
X. H. Sun,$^{2}$
W. Reich,$^{3}$
J. L. Han$^{1,4,5}$
\\
$^{1}$National Astronomical Observatories, Chinese Academy of Sciences,
20A Datun Road, Chaoyang District, Beijing 100101, China\\
$^{2}$School of Physics and Astronomy, Yunnan University, Kunming 650500,
PR China\\
$^{3}$Max-Planck-Institut f{\"u}r Radioastronomie, 53121 Bonn, Germany\\
$^{4}$CAS Key Laboratory of FAST, NAOC, Chinese Academy of Sciences,
Beijing 100101, China\\
$^{5}$School of Astronomy, University of Chinese Academy of Sciences,
Beijing 100049, China
}

\date{Accepted XXX. Received YYY; in original form ZZZ}

\pubyear{2023}

\begin{document}
\label{firstpage}
\pagerange{\pageref{firstpage}--\pageref{lastpage}}
\maketitle
\begin{abstract}
G213.0$-$0.6 is a faint extended source situated in the anti-center
region of the Galactic plane. It has been classified as a shell-type
supernova remnant (SNR) based on its shell-like morphology, steep
radio continuum spectrum, and high ratio of
[\ion{S}{II}]/H$\alpha$. With new optical emission line data of
H$\alpha$, [\ion{S}{II}], and [\ion{N}{II}] recently observed by the
Large Sky Area Multi-Object Fiber Spectroscopic Telescope, the ratios
of [\ion{S}{II}]/H$\alpha$ and [\ion{N}{II}]/H$\alpha$ are
re-assessed. The lower values than those previously reported put
G213.0$-$0.6 around the borderline of SNR-\ion{H}{II} region
classification. We decompose the steep-spectrum synchrotron and the
flat-spectrum thermal free-free emission in the area of G213.0$-$0.6
with multi-frequency radio continuum data. G213.0$-$0.6 is found to
show a flat spectrum, in conflict with the properties of a shell-type
SNR. Such a result is further confirmed by TT-plots made between the
863-MHz, 1.4-GHz, and 4.8-GHz data. Combining the evidence extracted
in both optical and radio continuum, we argue that G213.0$-$0.6 is
possibly not an SNR, but an \ion{H}{II} region instead. The $V_{LSR}$
pertaining to the H$\alpha$ filaments places G213.0$-$0.6
approximately 1.9~kpc away in the Perseus Arm.
\end{abstract}




\section{Introduction}
\label{sect:intro}
G213.0$-$0.6 is a faint extended source located in the anti-center
region of the Galactic plane. It has been proposed to be a shell-type
supernova remnant (SNR) by \citet{Reich03} based on its morphology and
the radio continuum spectral index of $\alpha = -0.40\pm0.15$ ($S_\nu
\sim \nu^{\alpha}$, with $S_\nu$ being the flux density at frequency
$\nu$), derived from a 30$\arcmin$-wide region across the source
between the Effelsberg 863-MHz and 2695-MHz data. An even earlier
study partially relevant to this target was conducted by
\citet{Bonsignori79}. They proposed the existence of another SNR
G211.7$-$1.1, which coincides with the nearby \ion{H}{II} region SH
2-284 \citep{Sharpless59} and includes the strong shell-like structure
of G213.0$-$0.6 (see Fig.~11 in their work). \citet{Stupar12} studied
G213.0$-$0.6 using optical observations. They confirmed it to be an
SNR based on the high [\ion{S}{II}]/H$\alpha$ ratio of 0.5 $-$ 1.1 for
some of its bright filaments. They also adjusted the name to
G213.3$-$0.4 according to a newly determined geometric center. In this
work, we adhere to the name of ``G213.0$-$0.6'' in line with SNR
catalogs \citep[e.g.][]{Green19} and other references.

Based on the spatial correlation with CO clouds, the distance of
G213.0$-$0.6 was suggested to be approximately 1~kpc
\citep{Su17}. This is supported by the distance assessments from
optical and dust extinction analysis \citep{Yu19, Zhao20}. The
apparent size $160\arcmin \times 140\arcmin$ of G213.0$-$0.6 therefore
corresponds to the physical extent of 46~pc $\times$ 40~pc. The
prominent \ion{H}{II} region SH 2-284, close to G213.0$-$0.6 in the
sky, has a distance of about 4~kpc \citep{Cusano11} or 4.5~kpc
\citep{Negueruela15}. This implies no physical connection between
G213.0$-$0.6 and SH 2-284.

Until now, studies that are focused on G213.0$-$0.6 are still very
limited, possibly because it is too faint and extended to be well
observed. In this paper, we aim to re-investigate the nature of
G213.0$-$0.6 by combining the recently observed optical spectral line
data from the Large Sky Area Multi-Object Fiber Spectroscopic
Telescope (LAMOST) and sensitive multi-frequency radio continuum data
which are public. In the following, we first introduce the LAMOST
optical emission line data and present the line ratios of
[\ion{S}{II}]/H$\alpha$ and [\ion{N}{II}]/H$\alpha$ newly derived from
the data in Sect.~\ref{sect:result1}. We show radio continuum images
derived from the component separation and the spectrum determined
through TT-plots in Sect.~\ref{sect:result2}. The morphology,
distance, and the possible ionizing source related to G213.0$-$0.6 are
discussed in Sect.~\ref{sect:dis}. The intriguing characteristics of
G213.0$-$0.6 are summarized in Sect.~\ref{sect:sum}.

\begin{figure*}
\centering
\includegraphics[angle=-90, width=0.9\textwidth]{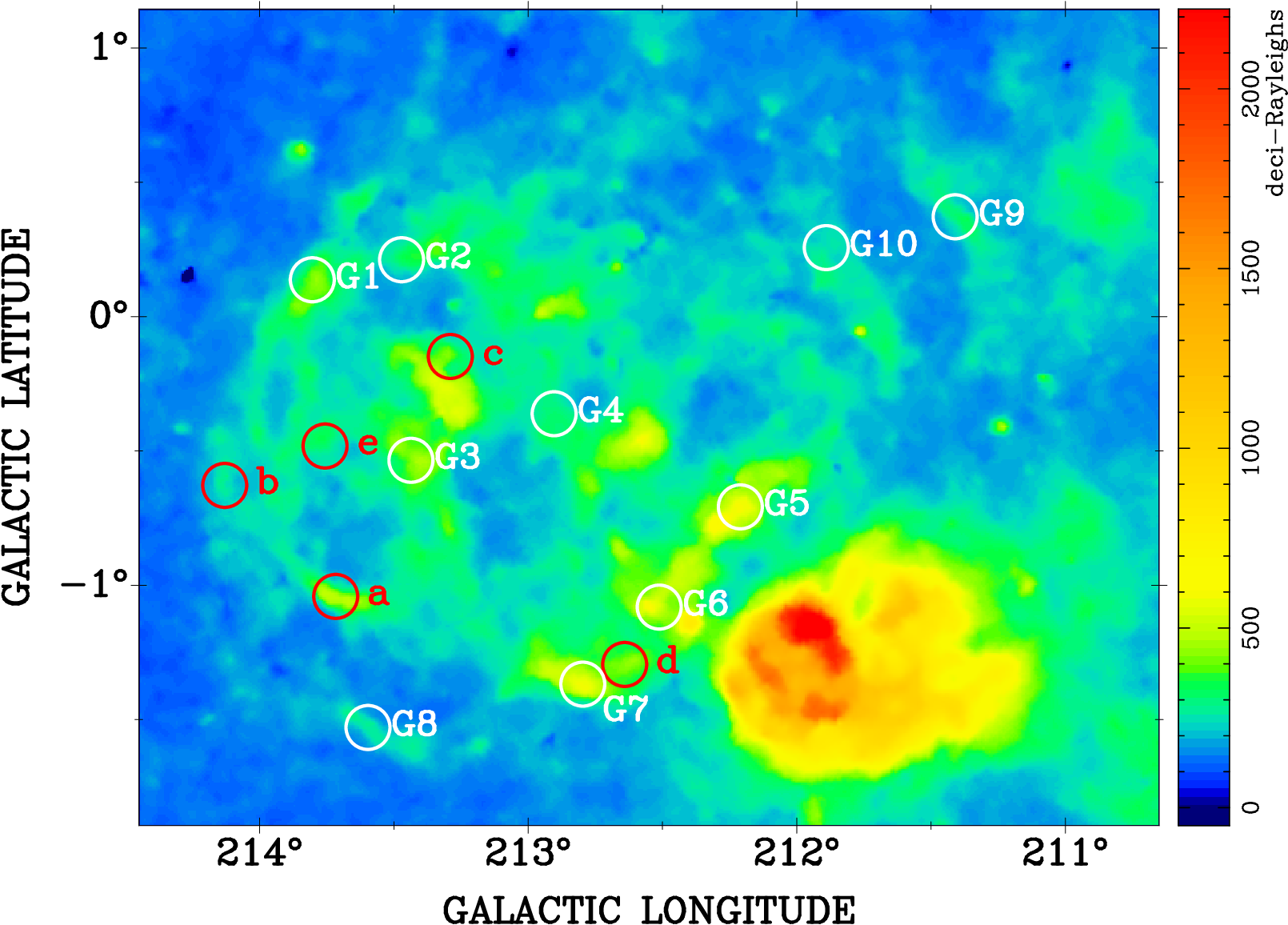}
\caption{H$\alpha$ emission in the area of G213.0$-$0.6 extracted from
  the Southern H$\alpha$ Sky Survey Atlas \citep{Gaustad01}. Five
  positions (red open circles labeled with ``a'' to ``e'') from
  \citet{Stupar12} and ten extra positions (white open circles
  starting with ``G'') selected in this work are used to derive the
  properties of the optical emission lines as listed in
  Table~\ref{Tab:tab1}.}
\label{fig:fg1}
\end{figure*}

\section{Optical emission lines}
\label{sect:result1}
H$\alpha$ images of G213.0$-$0.6 were shown in \citet{Stupar12} with
both the high-resolution data from the Anglo-Australian
Observatory/United Kingdom Schmidt Telescope (AAO/UKST) H$\alpha$
survey of the southern Galactic plane \citep{Parker05} and the
lower-resolution but high-sensitivity data from the Southern H$\alpha$
Sky Survey Atlas \citep[hereafter SHASSA,][]{Gaustad01}. Their
follow-up optical spectral observations, which were used for the
estimates of spectral line ratios, were made with the 1.9-m Radcliffe
telescope of the South African Astronomical Observatory. In this work,
we present the H$\alpha$ image of G213.0$-$0.6 using SHASSA data in
Fig.~\ref{fig:fg1}, because they better show the complete shell
structure of G213.0$-$0.6 as pointed out by \citet{Stupar12}. The
spectral data used for analysis are from LAMOST, a powerful instrument
for optical spectroscopy \citep{cui2012}, which is a four-meter
reflecting Schmidt telescope outfitted with 4\,000 fibers distributed
on its focal plane. We extract the processed data products from the
project of the LAMOST Medium-Resolution Spectral survey
\citep[MRS,][]{liu2020} of Galactic Nebulae
\citep[MRS-N,][]{Wu21}. They were observed in November 2021. The
wavelength range covered by the MRS-N comprises two segments: the red
arm spanning 6300\AA \ to 6800\AA \, and the blue arm encompassing
4950\AA \ to 5350\AA. Within the red arm, notable nebular emission
lines, i.e. H$\alpha$, [\ion{N} {II}]$\uplambda\uplambda$6548,6584 and
[\ion{S} {II}]$\uplambda\uplambda$6717,6731, have been recorded with a
relatively high spectral resolution (R$\sim$7500). \citet{Ren21}
introduced a method to re-calibrate wavelengths and achieve a radial
velocity measurement precision better than 1~km/s. \citet{Zhang21}
outlined a technique to subtract geo-coronal H$\alpha$ emission, that
successfully reduced the contamination level of the geo-coronal
H$\alpha$ line from 40\% to below 10\%. The detailed data processing
procedure can be referred to \citet{Wu22}.

\citet{Stupar12} analyzed spectral line ratios at five positions on
some of the prominent filaments of G213.0$-$0.6 and showed that all
the five [\ion{S} {II}]/H$\alpha$ ratios exceed 0.5, with a maximum of
1.1, favoring a shock-excited SNR origin. With the new LAMOST data, we
take measurements not only for the five positions (denoted as "a" to
"e" in Fig.~\ref{fig:fg1}) selected in \citet{Stupar12}, but also add
ten additional positions on other bright filaments, labeled from “G1”
to “G10” in Fig.~\ref{fig:fg1}. Because the separation of two adjacent
fibers of LAMOST is about 2$\arcmin$, the best spectra data within a
radius of 5 $\arcmin$ of each position with a signal-to-noise ratio
over 10 are chosen to make the average. The results are compiled in
Table~\ref{Tab:tab1}.
\begin{table*}
\begin{center}
  \caption{Properties of the optical spectral lines in the area of
    G213.0$-$0.6 observed by the LAMOST MRS-N. The 1st $-$ 3rd columns
    present the reference name and their coordinates marked in
    Fig.~\ref{fig:fg1}. The 4th column shows the $V_{LSR}$ measured
    for the H$\alpha$ line. The spectral line ratio of
    [\ion{N}{II}]/H$\alpha$ and [\ion{S}{II}]/H$\alpha$ measured in
    this work are listed in the 5th and 7th column. The values
    reported in \citet{Stupar12}, abbreviated as ``S12'' are presented
    in the 6th and 8th column for comparison. The ratio between the
    doublets of [\ion{S}{II}] and the estimates of the electron
    density are shown in the 9th and 10th column.}
  \label{Tab:tab1}
\begin{tabular}{cccccccccc}
\hline\hline\noalign{\smallskip}
\multicolumn{1}{c}{Reference} & RA (J2000) & Dec (J2000) &  $V_{LSR}$   & \multicolumn{1}{c}{[\ion{N}{II}]/H$\alpha$} & \multicolumn{1}{c}{[\ion{N}{II}]/H$\alpha$} & \multicolumn{1}{c}{[\ion{S}{II}]/H$\alpha$}  & \multicolumn{1}{c}{[\ion{S}{II}]/H$\alpha$} & \multicolumn{1}{c}{[\ion{S}{II}]}  & \multicolumn{1}{c}{Electron density} \\
& (h m s) & (d m s) &  (km/s)  &this work & S12 & this work & S12 & \multicolumn{1}{c}{(6717/6731\AA)}  & \multicolumn{1}{c}{(cm$^{-3}$)}\\
(1) & (2) & (3)     &   (4)    & (5)      & (6) & (7)       & (8) & (9)   & (10) \\
\hline\noalign{\smallskip}
a   & 06 49 09 & $-$01 10 30 & 15.6 & 0.45  & 0.76     & 0.44   & 0.6   & 1.27    & ~140 \\
b   & 06 51 22 & $-$01 21 17 & 13.6 & 0.40  & 0.59     & 0.36   & 0.8   & 1.49    & ~20 \\
c   & 06 51 33 & $-$00 23 03 & 22.7 & 0.42  & 0.50     & 0.41   & 0.5   & 1.59    & ~10 \\
d   & 06 46 16 & $-$00 20 07 & 22.6 & 0.39  & 0.58     & 0.51   & 1.1   & 1.45    & ~30 \\
e   & 06 51 13 & $-$00 57 03 & 12.8 & 0.45  & 0.88     & 0.38   & 1.0   & 1.50    & ~19 \\
G1  & 06 53 30 & $-$00 42 58 & 18.1 & 0.42  &  $-$     & 0.45   & $-$   & 1.41    & ~50 \\
G2  & 06 53 09 & $-$00 23 11 & 12.5 & 0.47  &  $-$     & 0.48   & $-$   & 1.50    & ~19 \\
G3  & 06 50 26 & $-$00 41 51 & 25.1 & 0.49  &  $-$     & 0.38   & $-$   & 1.42    & ~44 \\
G4  & 06 50 05 & $-$00 08 30 & 17.7 & 0.32  &  $-$     & 0.32   & $-$   & 1.29    & ~120 \\
G5  & 06 47 36 & $+$00 18 56 & 27.1 & 0.37  &  $-$     & 0.37   & $-$   & 1.43    & ~30 \\
G6  & 06 46 48 & $-$00 07 22 & 30.6 & 0.35  &  $-$     & 0.41   & $-$   & 1.46    & ~30 \\
G7  & 06 46 18 & $-$00 30 24 & 22.4 & 0.36  &  $-$     & 0.42   & $-$   & 1.37    & ~70 \\
G8  & 06 47 11 & $-$01 17 29 & 6.1  & 0.29  &  $-$     & 0.43   & $-$   & 1.35    & ~90 \\
G9  & 06 49 58 & $+$01 31 12 & 26.8 & 0.43  &  $-$     & 0.55   & $-$   & 1.51    & ~10 \\
G10 & 06 50 26 & $+$01 02 35 & 24.7 & 0.53  &  $-$     & 0.59   & $-$   & 1.30    & ~110 \\
\noalign{\smallskip}\hline
\end{tabular}
\end{center}
\end{table*}

The LAMOST measurements (see Table~\ref{Tab:tab1}) show that the
[\ion{S}{II}]/H$\alpha$ and [\ion{N}{II}]/H$\alpha$ ratios for the
five common positions as in \citet{Stupar12} are mostly below
0.5. Some are nearly half of the values previously reported. For the
rest ten positions newly selected in this work, the results are
consistent with the new values for the five common positions. The
average values of [\ion{S}{II}]/H$\alpha$ and [\ion{N}{II}]/H$\alpha$
for all the 15 positions are about 0.43 and 0.41, respectively. The
criterion value of [\ion{S}{II}]/H$\alpha$ ratio to distinguish SNRs
and \ion{H}{II} regions has long been discussed. Besides the value of
0.5 as quoted in e.g. \citet{Stupar12}, 0.4 and 2/3 were also adopted
\citep[see][and references therein]{Fesen85}. To solve the ambiguity
when 0.4 < [\ion{S}{II}]/H$\alpha$ < 2/3, \citet{Fesen85} suggested to
use the line ratios of [\ion{O}{I}]/H$\beta$ and
[\ion{O}{II}]/H$\beta$ to aid the classification. In a recent work,
\citet{Kopsacheili20} explored more efficient 3D and 2D diagnostic
tools by using [\ion{O}{I}]/H$\alpha$ $-$ [\ion{O}{II}]/H$\beta$ $-$
[\ion{O}{III}]/H$\beta$ and [\ion{O}{I}]/H$\alpha$ $-$
[\ion{O}{III}]/H$\beta$ with a low confusion in
classification. However, our LAMOST data do not cover
[\ion{O}{II}]($\mathrm{\uplambda\uplambda 3727,3729}$) or
H$\beta$($\uplambda4862$). The [\ion{O}{I}]($\uplambda$ 6300) line is
contaminated by the skylight. In particular, we notice that even if
the criterion value of [\ion{S}{II}]/H$\alpha$ is set to 0.4, one
still cannot completely rule out some \ion{H}{II} regions that could
be mistaken as SNRs \citep[see][in their
  Fig.~5]{Kopsacheili20}. Different values of [\ion{S}{II}]/H$\alpha$
both greater and less than 0.4 are also seen within the same
\ion{H}{II} region \citep[e.g.,][]{Li22}. Another method utilizing
optical emission line ratios, considering both
log([\ion{S}{II}]/H$\alpha$) and log([\ion{N}{II}]/H$\alpha$), has
been employed by e.g. \citet{Kniazev08, Lagrois12} and
\citet{Sabin13}. The average values from the LAMOST measurements
correspond to log([\ion{S}{II}]/H$\alpha$) $\sim$ $-$0.36 and
log([\ion{N}{II}]/H$\alpha$) $\sim$ $-$0.39, also placing G213.0$-$0.6
in the borderline area between ``\ion{H}{II} region'' and ``SNR''
\citep[see Fig.~4 in][]{Kniazev08}. Thus, the current LAMOST data are
not sufficient to make a definite classification, but could raise a
question about the identity of G213.0$-$0.6 as being an SNR.

\section{Thermal emission nature revealed by Radio 
continuum}
\label{sect:result2}
Radio continuum emission and the corresponding spectral index provide
another useful tool in distinguishing between SNRs and \ion{H}{II}
regions \citep[e.g.][]{Foster06, Gao11}. Shell-type SNRs have
steep-spectrum synchrotron emission \citep[e.g.][]{Dubner15}, while
the optical-thin free-free emission from \ion{H}{II} regions shows a
flat spectrum ($\alpha \sim -0.1$, $S_{\nu} \sim \nu^{\alpha}$). With
publicly available radio continuum data, we investigate the radio
continuum emission properties of G213.0$-$0.6 via two methods in the
following.
\subsection{Component separation}
By using multi-frequency radio continuum data, \citet{Paladini05}
developed a method to decompose the mixed signal from the Galaxy into
non-thermal synchrotron emission stemming from relativistic electrons
spiraling in the magnetic field and thermal free-free emission from
ionized gas. Based on this method, \citet{Sun11a} separated the two
emission components in the inner Galactic plane area with $10\degr
\leq \ell \leq 60\degr$, and $|b| \leqslant 4\degr$, while
\citet{Xu13} scoured the Cygnus-X complex for new SNRs by singling out
synchrotron emission from strong confusing thermal emission. These
successful applications make the method useful in fast recognition of
the emission nature of radio sources in vast sky areas.

\begin{figure*}
\centering
\includegraphics[angle=-90,width=0.48\textwidth]{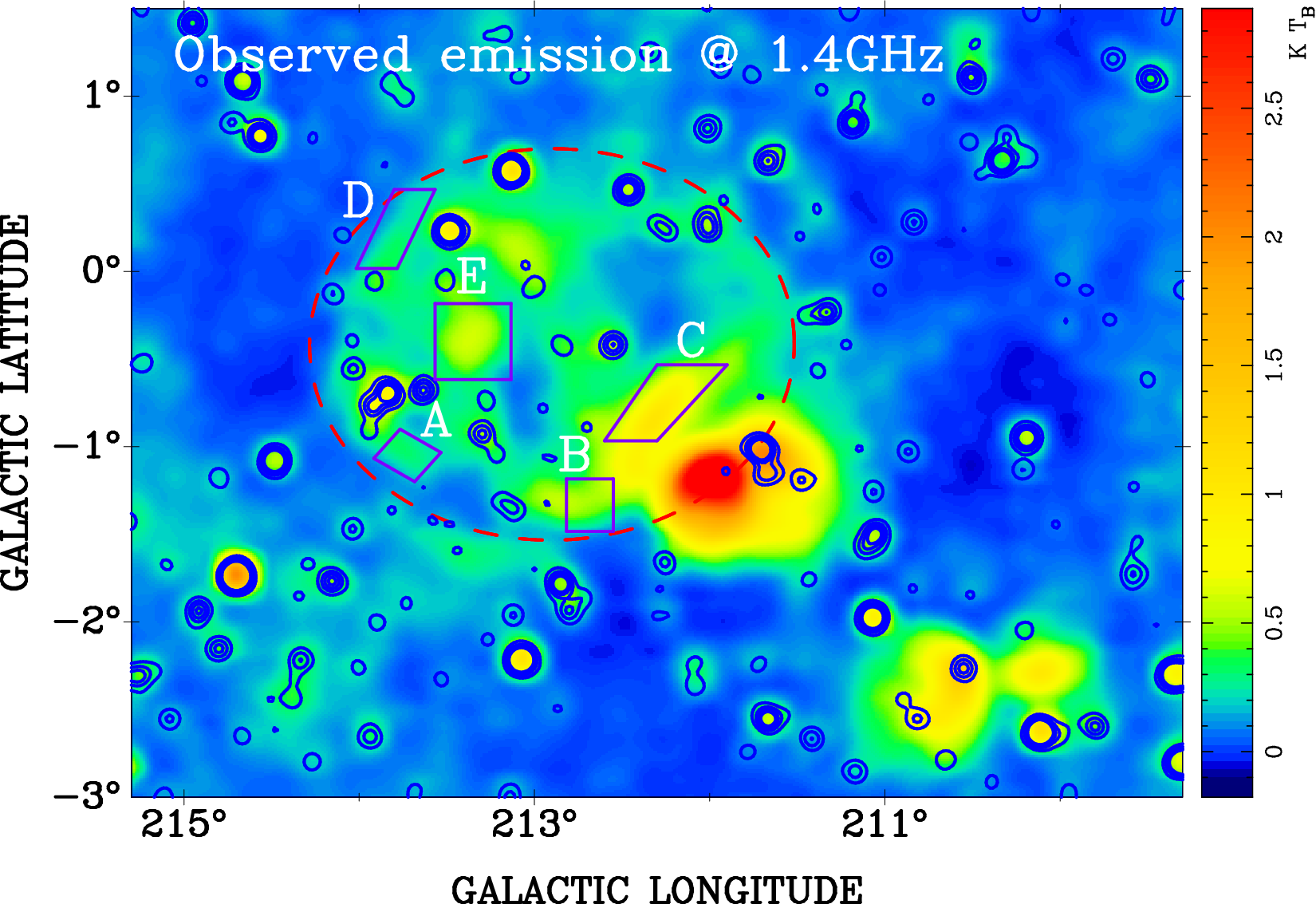}
\includegraphics[angle=-90,width=0.48\textwidth]{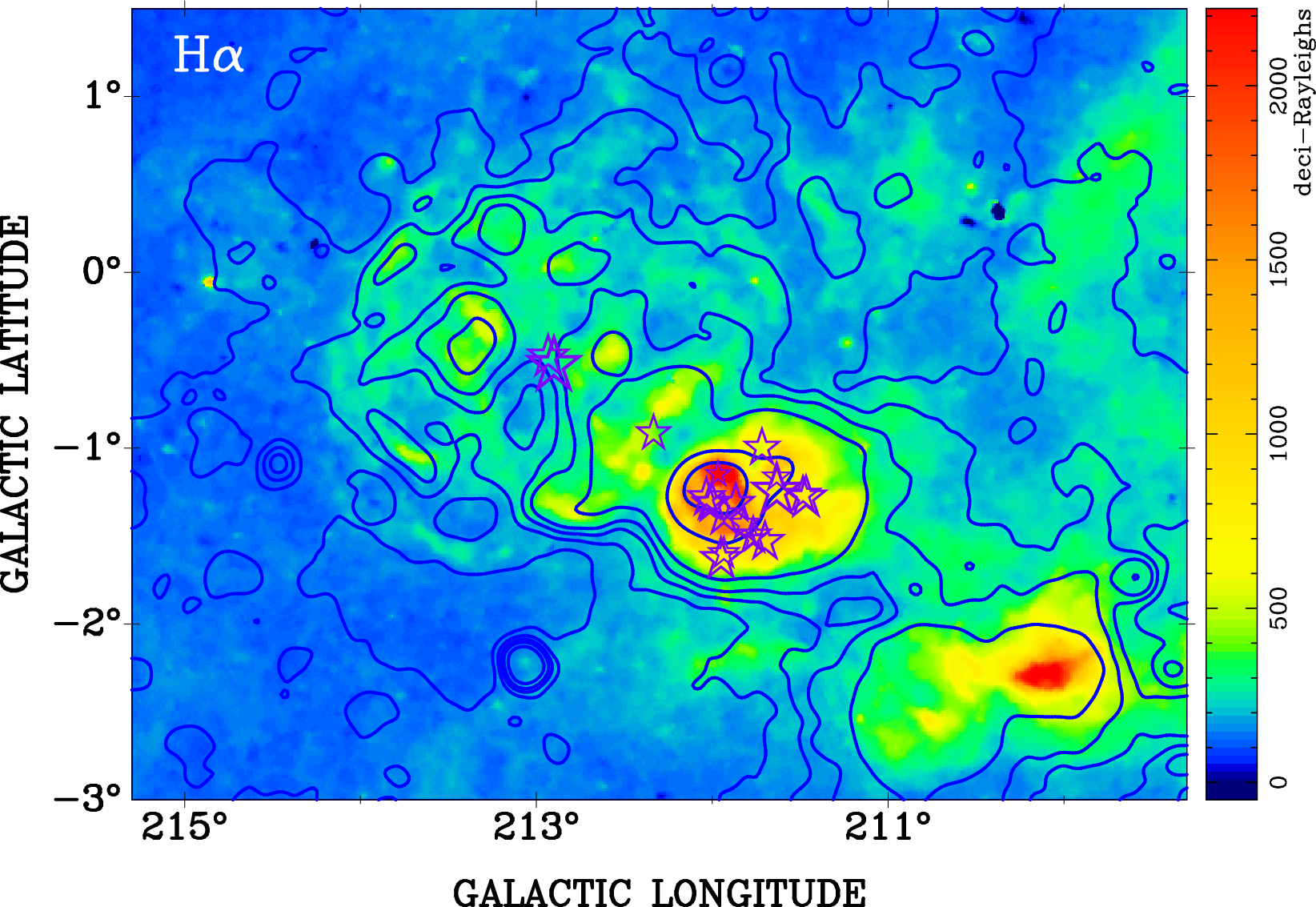}\\
\includegraphics[angle=-90,width=0.48\textwidth]{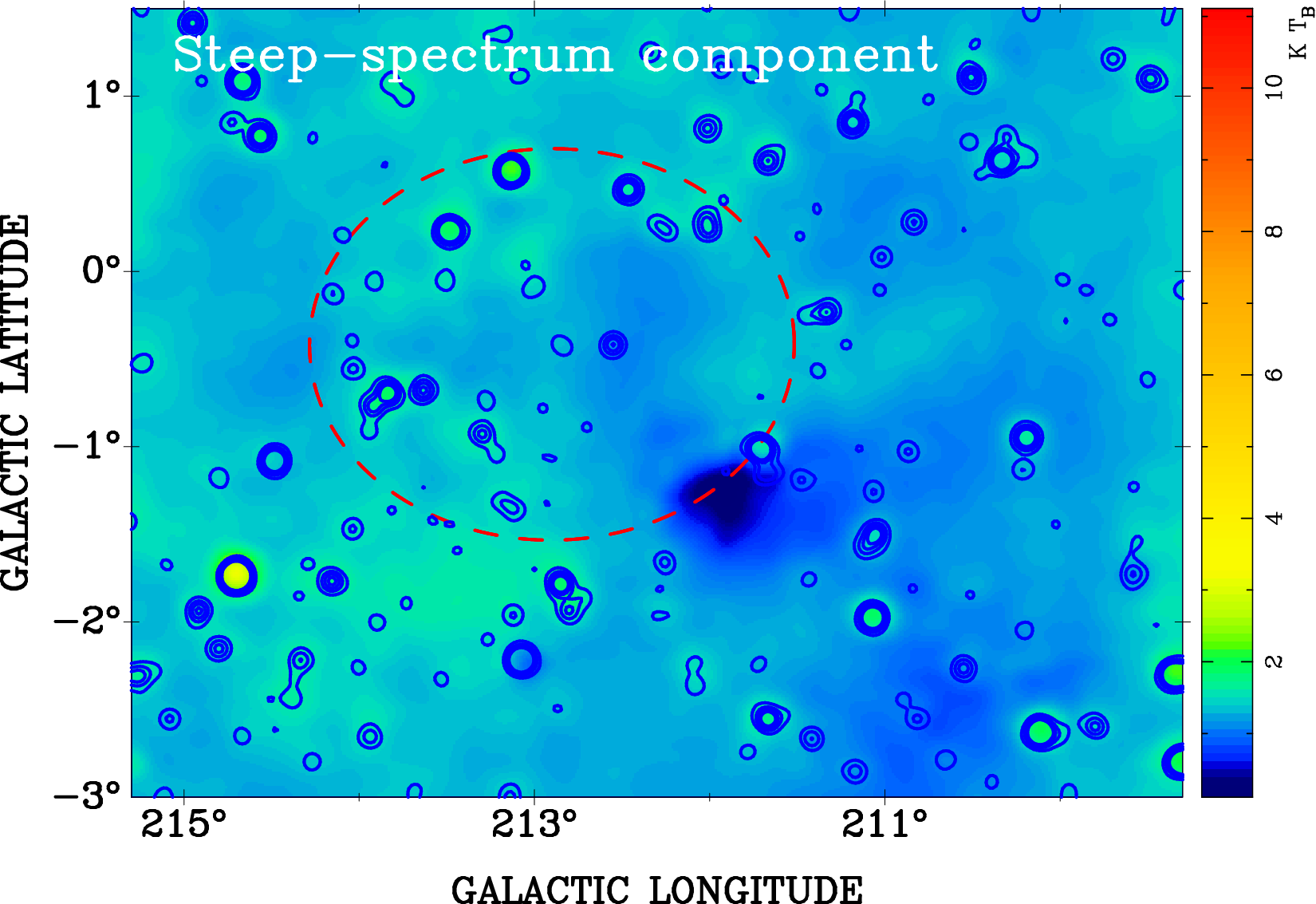}
\includegraphics[angle=-90,width=0.48\textwidth]{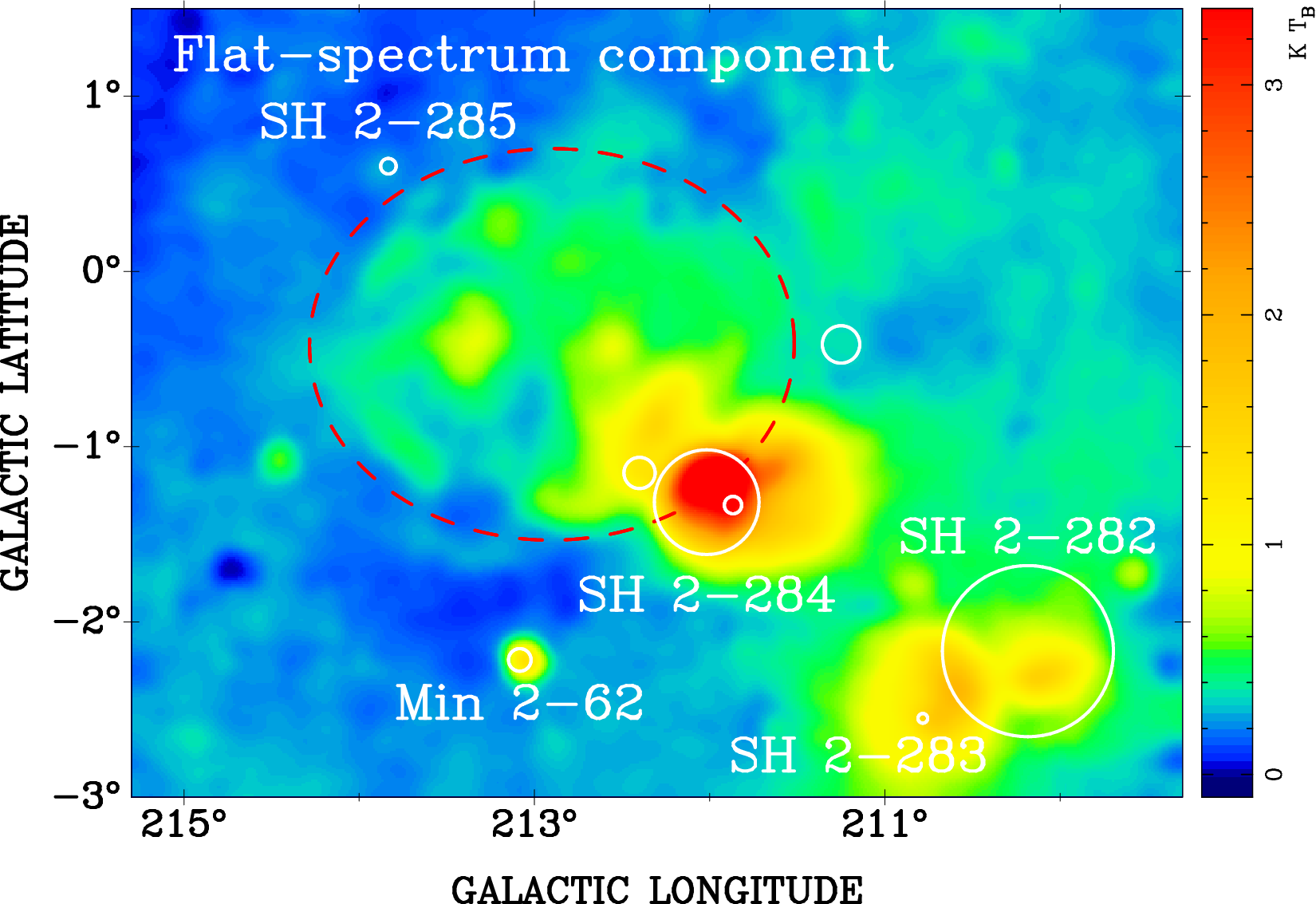}\\
\caption{{\it Upper-left panel:} Effelsberg 1.4-GHz total-intensity
  (source component) image for the area of G213.0$-$0.6 at a
  resolution of 9$\farcm$4. The large-scale diffuse Galactic
  background emission is not included. The ellipse enclosed by the red
  dashed line outlines G213.0$-$0.6. The blue contours indicate strong
  point-like sources in the NVSS \citep{Condon98} smoothed to
  6$\arcmin$.  The polygon areas illustrated by the purple lines and
  marked as ``A'' to ``E'' are used for TT-plots to derive the spectra
  of G213.0$-$0.6.  {\it Lower-left} and {\it lower-right panels:}
  images of the separated steep-spectrum and the flat-spectrum
  components smoothed to a resolution of 11$\arcmin$, including the
  Galactic large-scale emission.  The WISE \ion{H}{II} regions
  \citep{Anderson14} are illustrated by white circles in the {\it
    lower-right panel} with sizes representing their angular
  extent. The circles with names aside are \ion{H}{II} regions which
  have also been identified in previous optical observations
  \citep{Sharpless59, Frew13}. In the {\it upper-right panel},
  H$\alpha$ emission extracted from SHASSA \citep{Gaustad01} is
  presented overlaid by contours of the separated flat-spectrum radio
  emission. The purple stars are OB-type stars observed by Gaia and
  extracted from \citet{Xu21}. The sizes of the stars are scaled
  according to their distance as listed in Table~\ref{Tab:taba1}.}
\label{fig:fg2}
\end{figure*}

We improve the work of \citet{Xu13} by considering the steepening of
the synchrotron spectrum at higher frequencies and apply the algorithm
to the anti-center region of the Galactic plane. The result indicates
a strong tendency that G213.0$-$0.6 is a thermal-emitting source
rather than an SNR. We briefly explain the procedure and show the
results in below.

\citet{Paladini05} derived the thermal emission fraction via:
\begin{gather}
\label{eq:eq1}
f_{\rm th,\nu_{1}} = \frac{T_{\rm ff,\nu_{1}}}{T_{\rm gal,\nu_{1}}} =
\frac{T_{\rm ff,\nu_{1}}}{T_{\rm ff,\nu_{1}} + T_{\rm syn,\nu_{1}}} =
\frac{1 - (\nu_{2}/\nu_{1})^{\rm \beta_{gal} - \beta_{syn}}}{\rm 1 -
  (\nu_{2}/\nu_{1})^{\beta_{ff} - \beta_{syn}}}
\end{gather}
Here, $f_{\rm th,\nu_{1}}$ is the fraction of the thermal component at
frequency $\nu_{1}$, $T_{\rm gal}$ refers to the total Galactic
emission, i.e. the sum of non-thermal emission $T_{\rm syn}$ and the
thermal emission $T_{\rm ff}$. The parameters $\beta_{\rm gal}$,
$\beta_{\rm syn}$, and $\beta_{\rm ff}$ are the brightness-temperature
spectral indices of the Galactic, synchrotron and free-free emission,
respectively (T$_\nu \sim \nu^{\beta}$, $\beta = \alpha - 2$). The
value of $\beta_{\rm gal}$ can be conveniently calculated from the
observational data at a second frequency $\nu_{2}$ through $\beta_{\rm
  gal} = {\rm log}(T_{\rm gal,\nu_1}/T_{\rm gal,\nu_2}) / {\rm
  log}({\rm \nu_1}/{\rm \nu_2})$, and $\beta_{\rm ff}$ is usually set
to a fixed value of $-$2.1 \citep[e.g.][]{Paladini05} or alternatively
$-$2.15 \citep[e.g.][]{Bennett13}. Here, we adopt $\beta_{\rm ff} =
-2.15$. $\beta_{\rm syn}$ is the only unknown in the expression.

We use four data sets for component separation. These are the 408-MHz
all-sky survey data \citep{Haslam82}, which were de-striped by
\citet{Remazeilles15}, the Effelsberg 1408-MHz survey data
\citep{Reich97}, the 9-yr WMAP K-band (22.8-GHz) data
\citep{Bennett13}, and the Urumqi 4.8-GHz survey data of
\citet{Gao10}. As in \citet{Xu13}, after corrected for the offset
values such as errors in zero levels, the cosmic microwave background,
and the unresolved extra-galactic sources, the data of 408~MHz
v.s. 1408~MHz and the data of 1408~MHz v.s. 22.8~GHz at a common
angular resolution of 1$\degr$ are split into two groups. Because the
thermal emission at 1.4~GHz calculated from the first data pair should
be the same as that obtained from the second data pair, one can find
the best-fit value of $\beta_{\rm syn}$ for each pixel to balance the
two sides and then obtain the total amount of flat-spectrum and
steep-spectrum emission components at 1.4~GHz. Meanwhile, one can also
get the Galactic emission template at any frequency $\nu$ following
$T_{\rm gal,\nu} = T_{\rm syn, 1.4~GHz} \cdot (\nu/1.4)^{\beta_{\rm
    syn}} + T_{\rm ff, 1.4~GHz} \cdot (\nu/1.4)^{\beta_{\rm
    ff}}$. This template is used to compensate the missing large-scale
structure in the Urumqi 4.8-GHz total-intensity data. Subsequently, by
using the 9$\farcm$5 resolution data at 4.8~GHz and 1.4~GHz, together
with the synchrotron spectral indices found above, we could decompose
the Galactic emission via Eq.~\ref{eq:eq1}.

In practice, the synchrotron spectrum is not a perfect single-power
law but steepens as the observing frequency increases.
\citet{Jaffe11} found $-2.8 < \beta_{\rm syn} < -2.74$ from 408~MHz to
2.3~GHz and $-2.98 < \beta_{\rm syn} < -2.91$ from 2.3~GHz to
23~GHz. The model proposed by \citet{Kogut12} showed consistent
values, also with a difference of $\Delta\beta_{\rm syn} \sim 0.2$
between the low- and high-frequency data pairs. Based on this
information, we treat the synchrotron spectral index $\beta_{\rm syn}$
as two separate variables for the two groups rather than maintaining
the same value as in \citet{Xu13}. We seek for the best fit on the two
sides by constraining the difference of $\beta_{\rm syn}$ between the
two data pairs.

The decomposition results made for the 1.4-GHz data after considering
the synchrotron spectrum steepening are shown in
Fig.~\ref{fig:fg2}. The uncertainty, including those in the input data
sets and by the error propagation, is typically $\sim$15\%, lower than
those in \citet{Xu13}. At a first glance, many compact radio sources
are seen overlaid on the non-thermal background, however, almost all
the prominent extended radio sources are presented in the
flat-spectrum component image. They are mostly known \ion{H}{II}
regions, i.e. SH 2-282 to 285 and Min 2-62 (white circles in the {\it
  lower-right panel} of Fig.~\ref{fig:fg2}). Notably, G213.0$-$0.6
also appears to have a flat spectrum, not only the filaments that can
be resolved in the northeast, southeast, south, but also the strong
and thick shell structure in its southwest, where the thermal emission
fraction can reach $\sim$60\%. These separated radio continuum
flat-spectrum structures correlate well with the H$\alpha$ emission
\citep{Gaustad01} of G213.0$-$0.6 as shown in the {\it upper-right
  panel} of the Fig.~\ref{fig:fg2}.
\begin{figure*}
\centering
\includegraphics[angle=-90,width=0.32\textwidth]{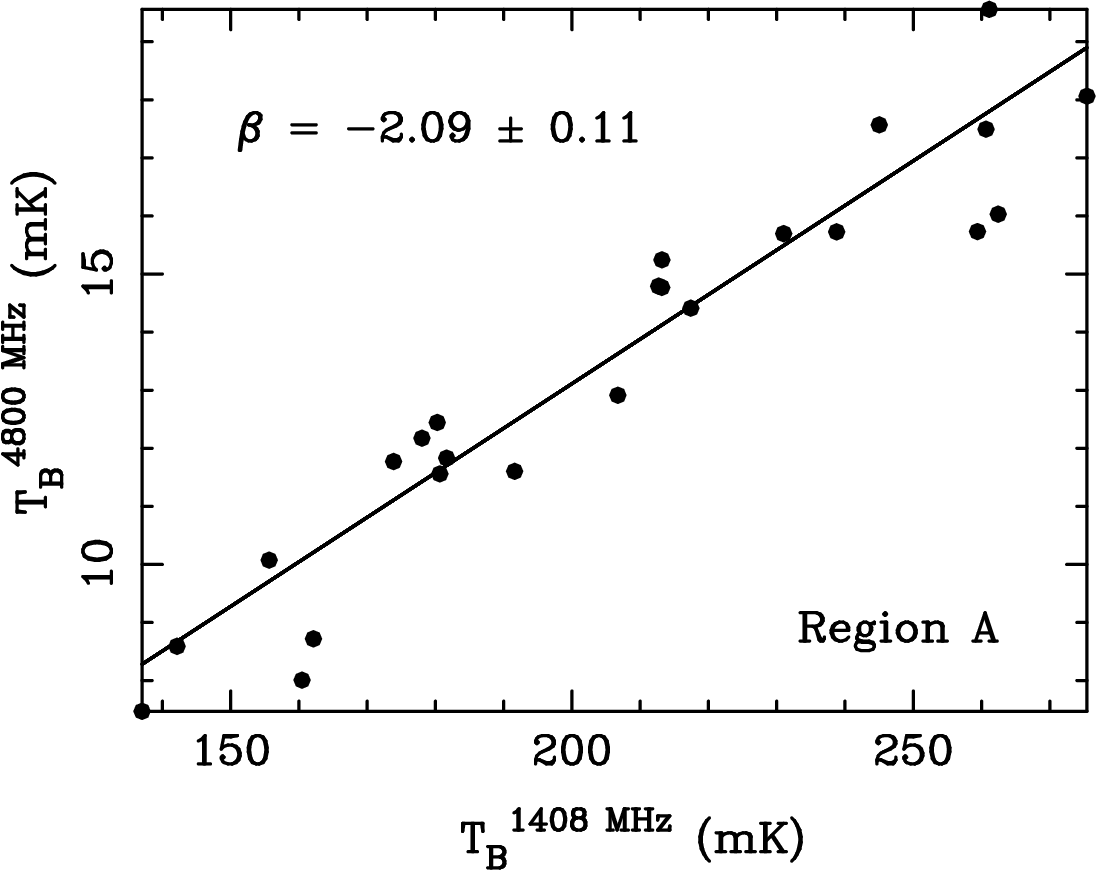}
\includegraphics[angle=-90,width=0.32\textwidth]{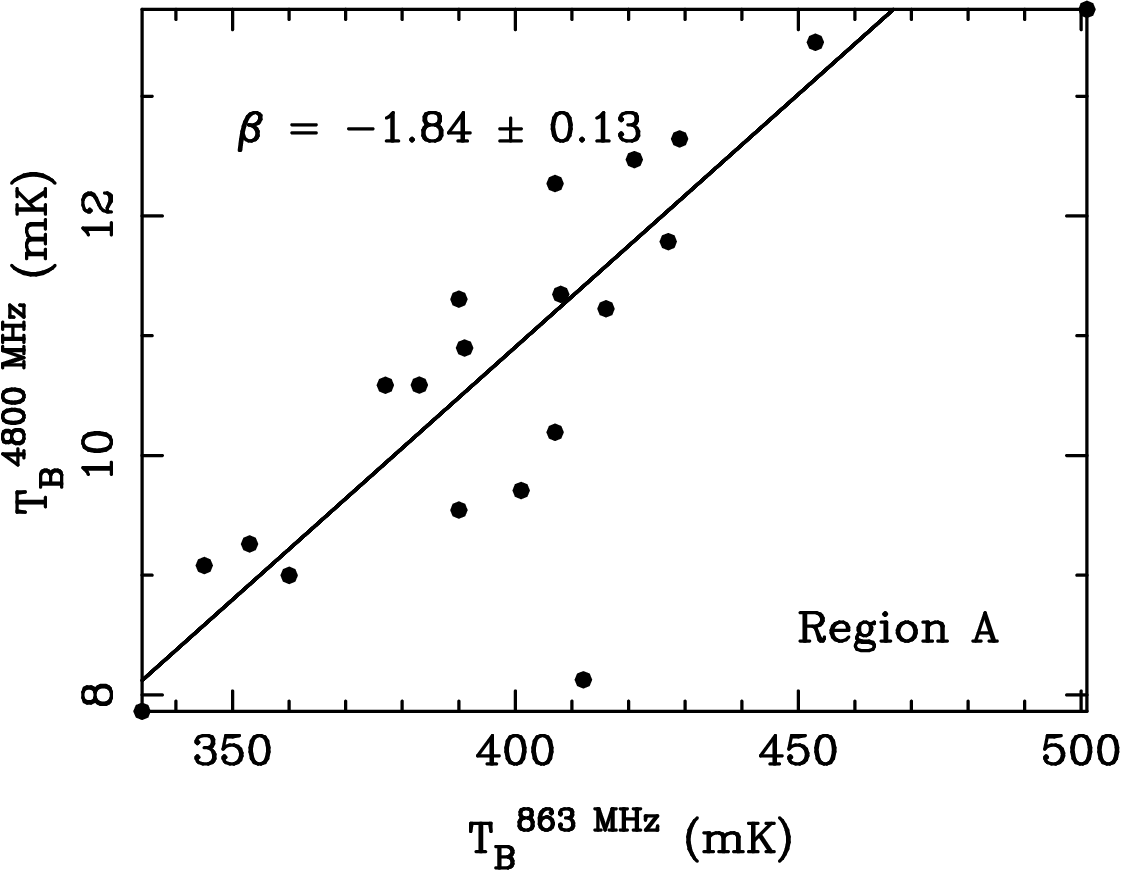}\\
\includegraphics[angle=-90,width=0.32\textwidth]{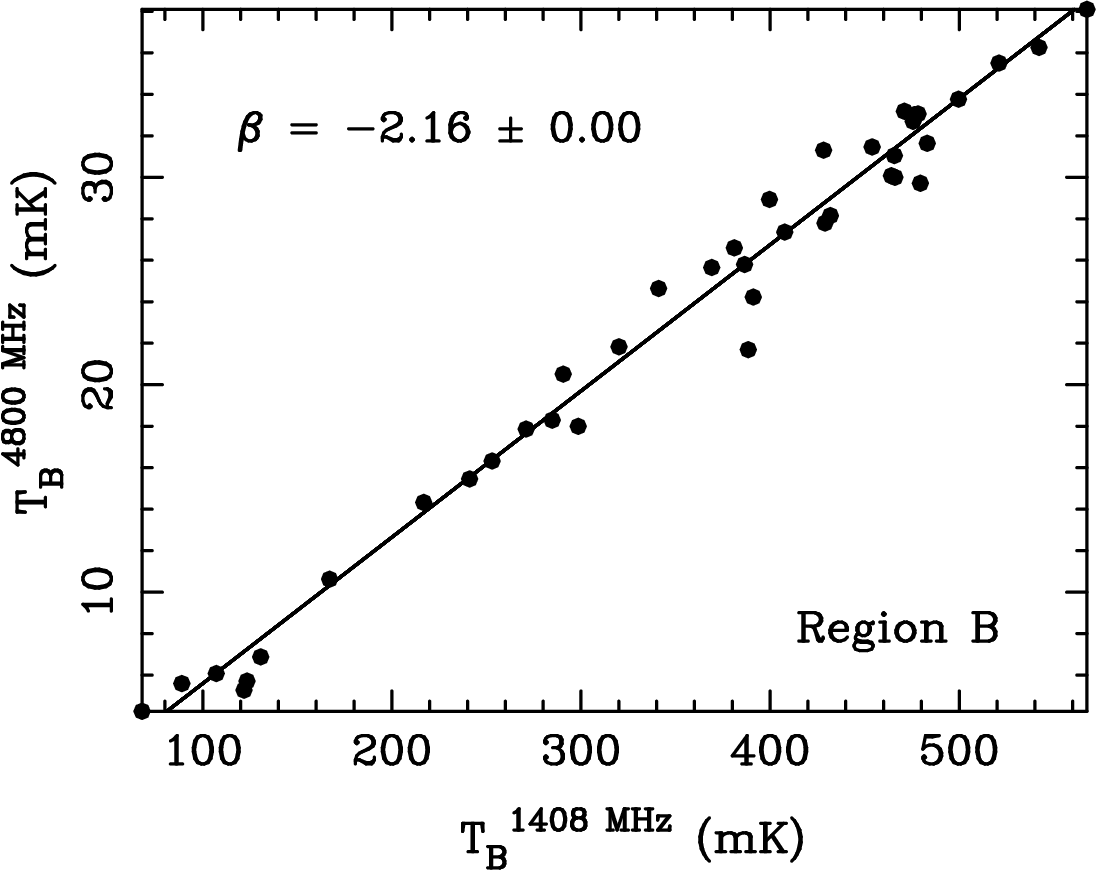}
\includegraphics[angle=-90,width=0.32\textwidth]{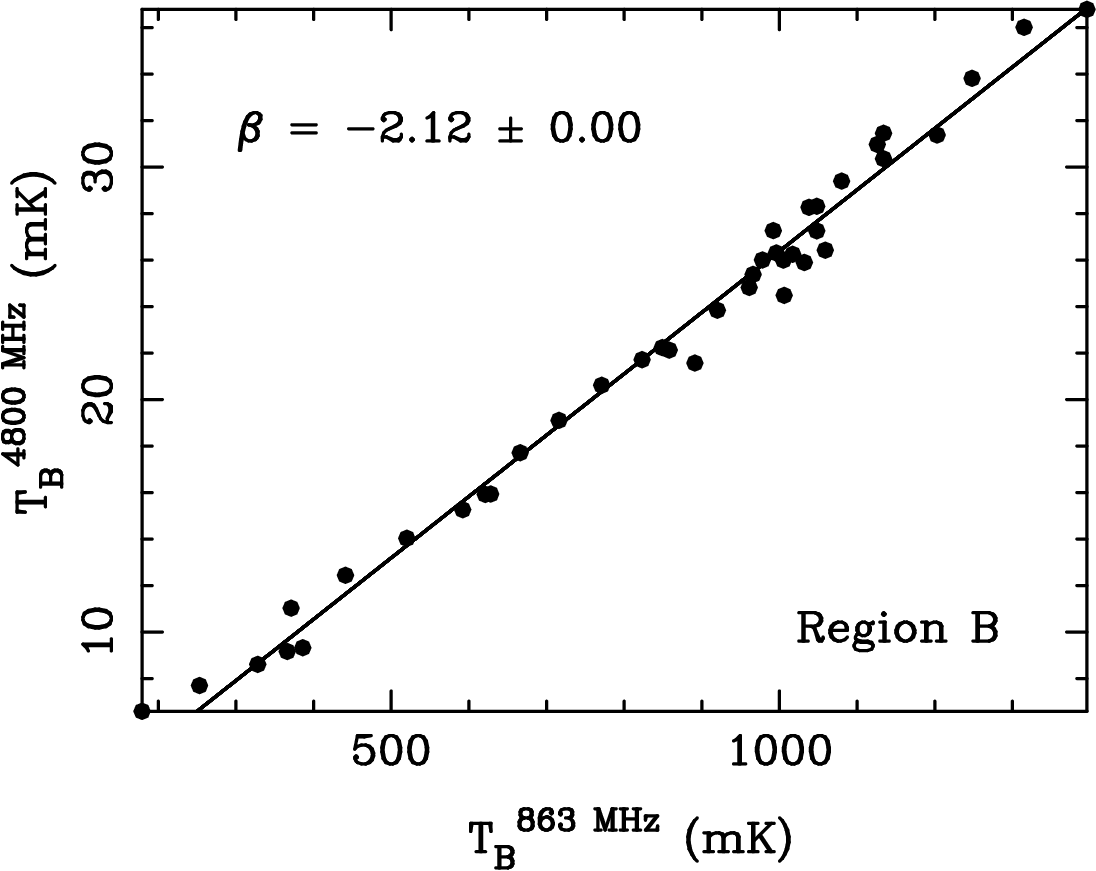}\\
\includegraphics[angle=-90,width=0.32\textwidth]{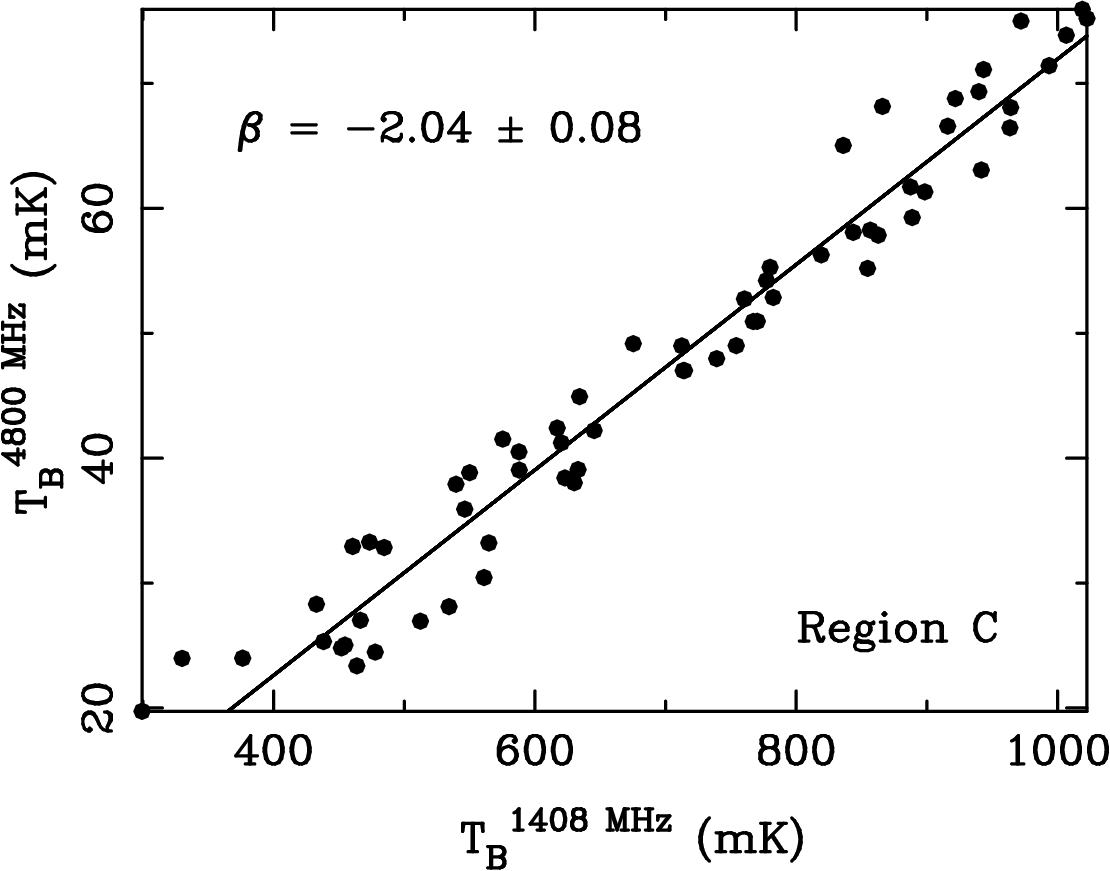}
\includegraphics[angle=-90,width=0.32\textwidth]{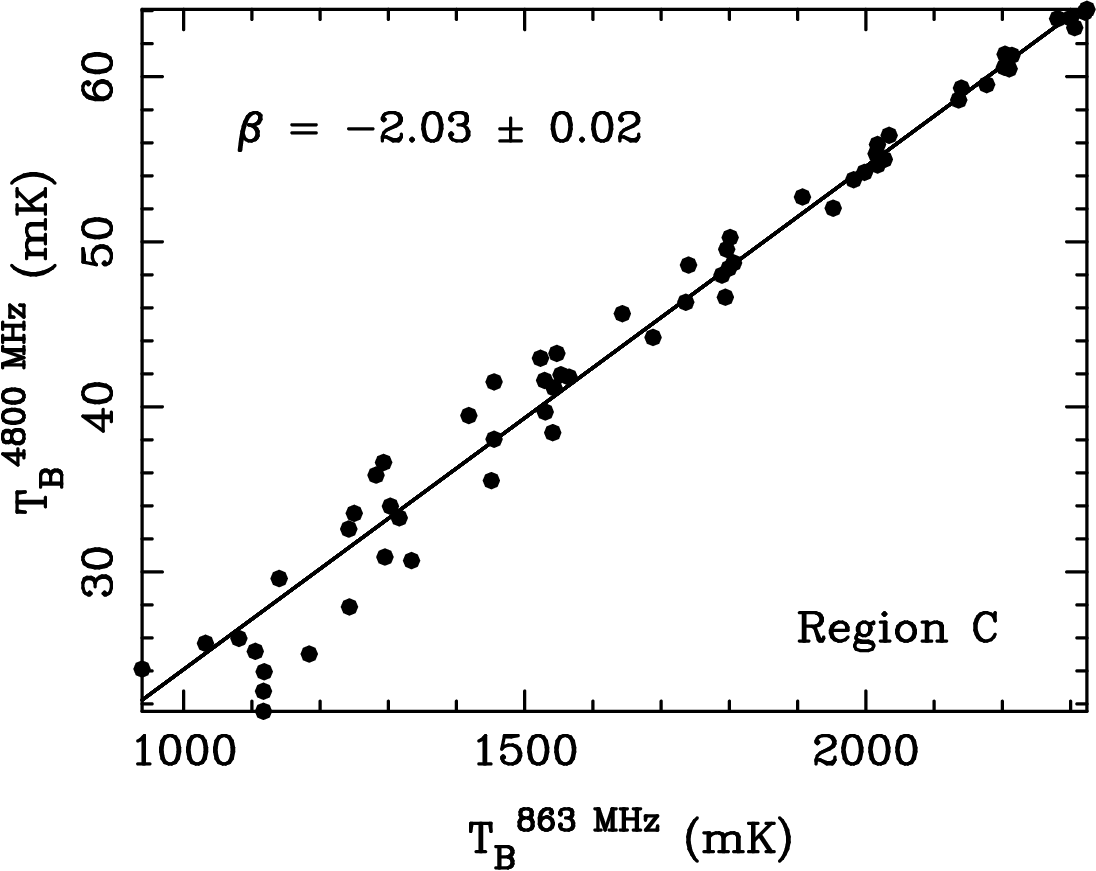}\\
\includegraphics[angle=-90,width=0.32\textwidth]{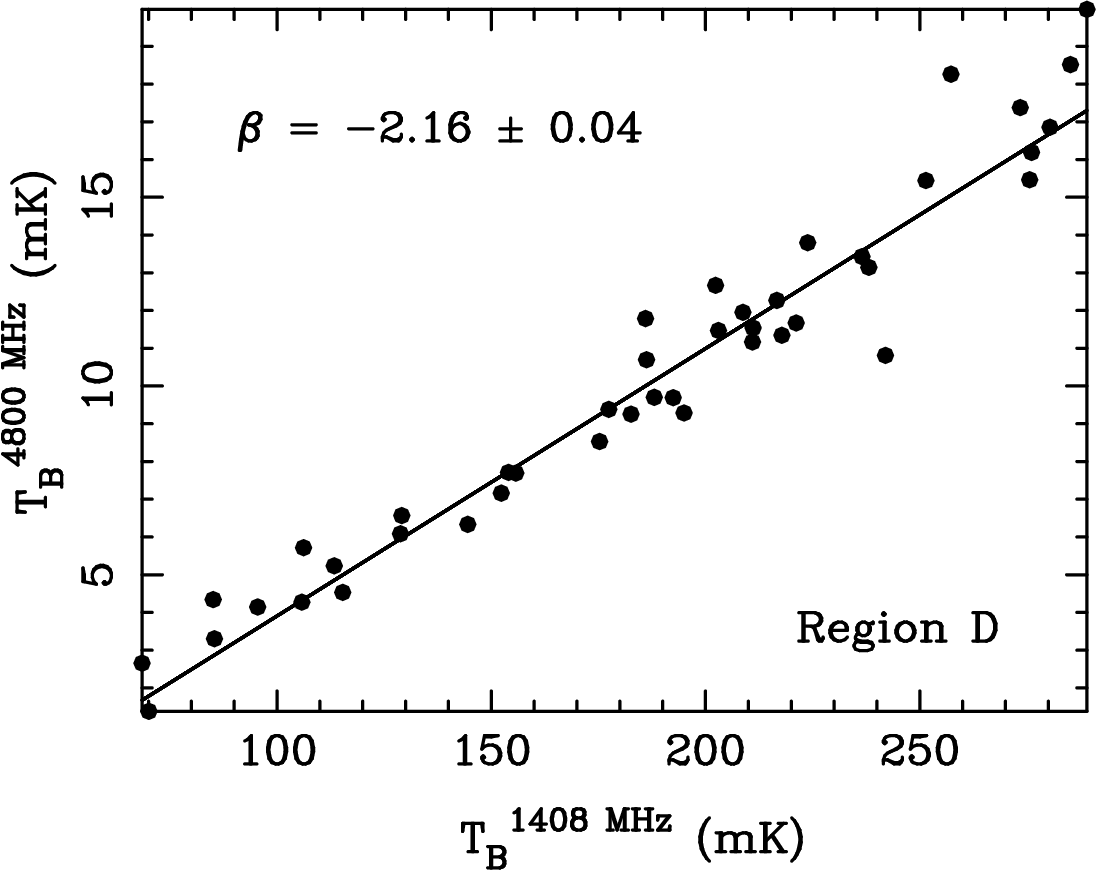}
\includegraphics[angle=-90,width=0.32\textwidth]{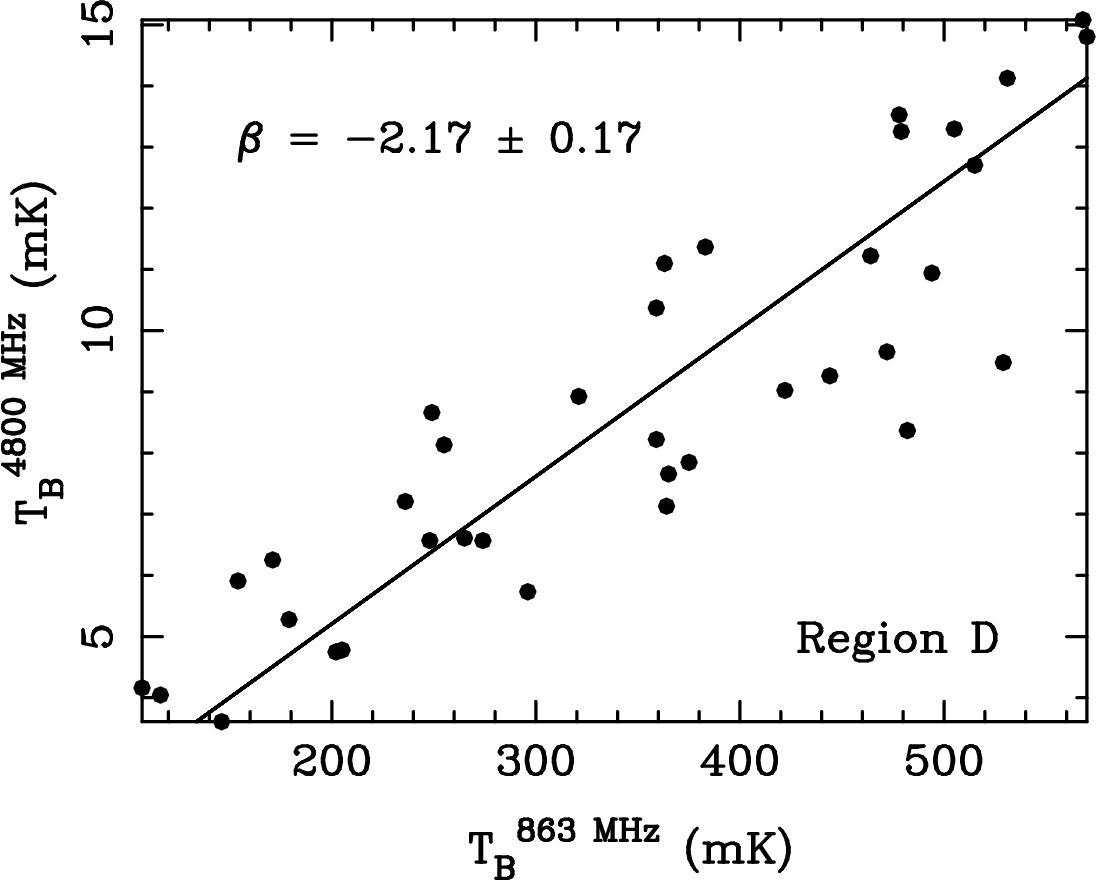}\\
\includegraphics[angle=-90,width=0.32\textwidth]{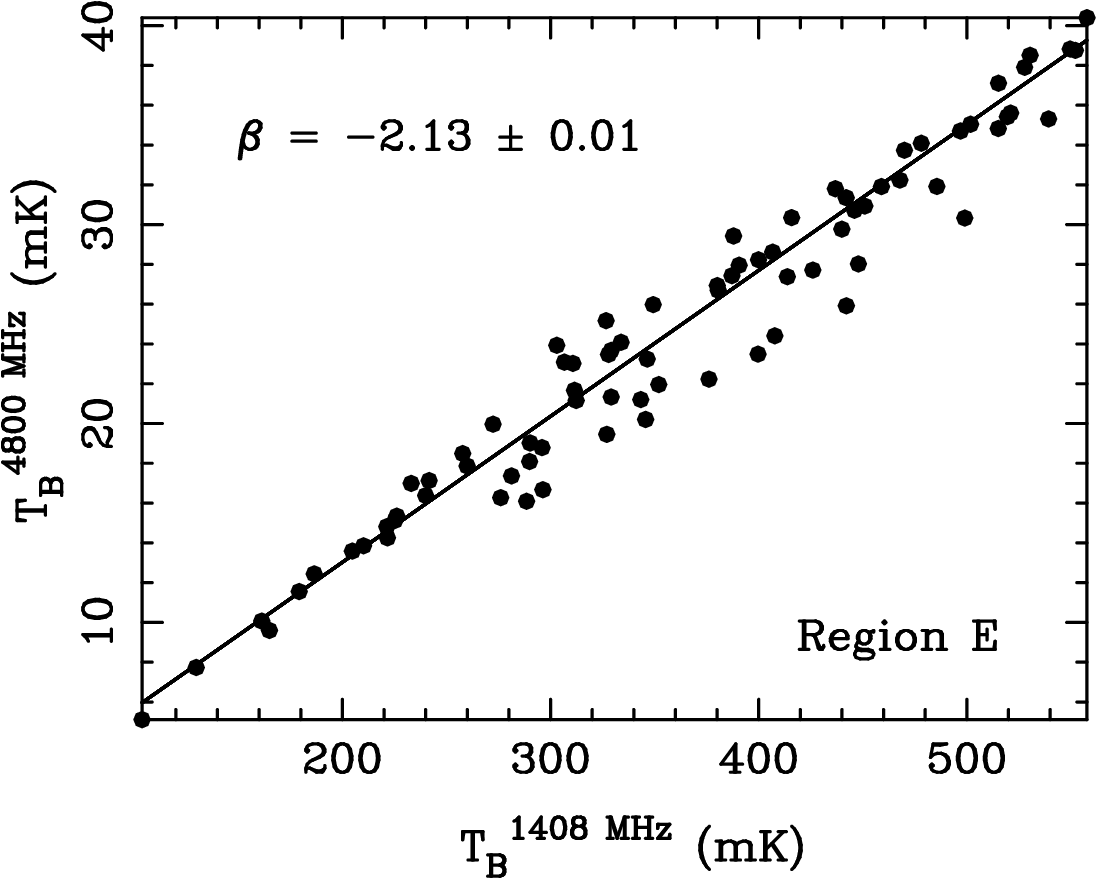}
\includegraphics[angle=-90,width=0.32\textwidth]{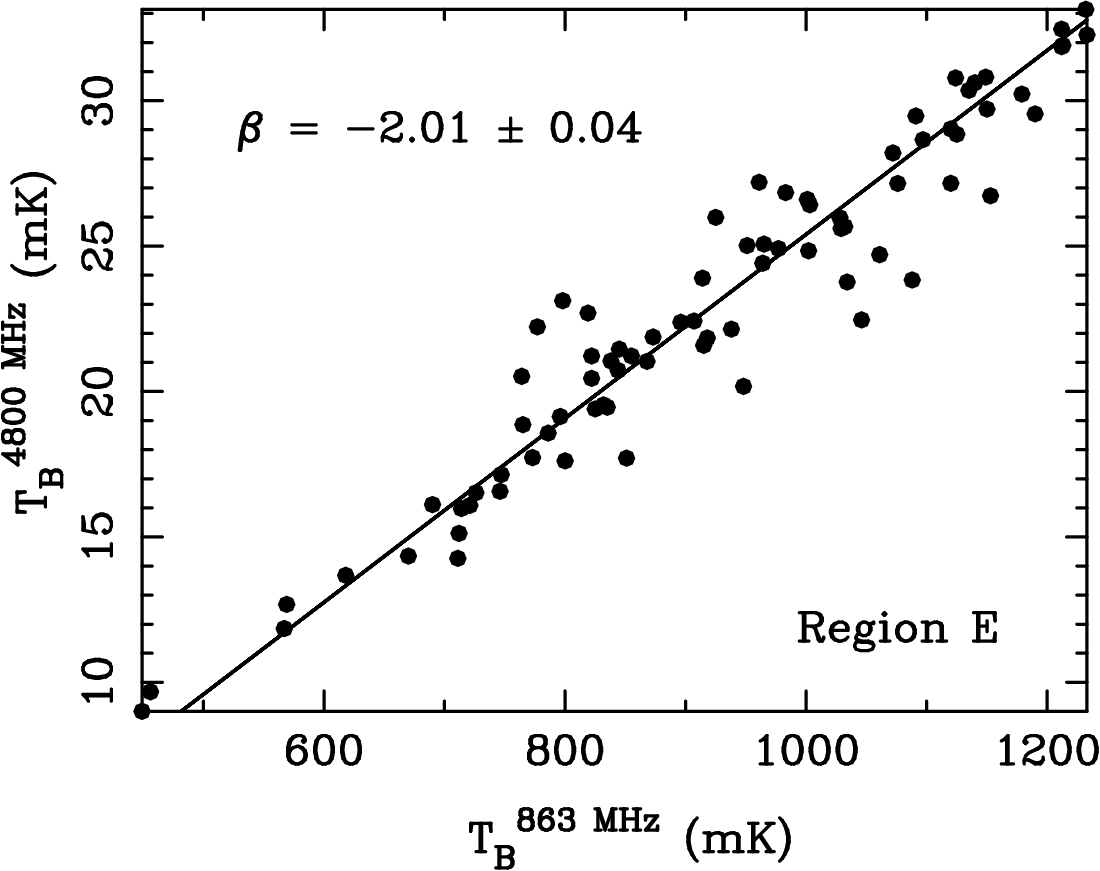}\\
\caption{TT-plots made for the spectral indices by using the data
  between 1.4~GHz (Effelsberg) and 4.8~GHz (Urumqi) at a common
  angular resolution of 9$\farcm$5 ({\it left panels}) and the data
  between 863~MHz (Effelsberg) and 4.8~GHz (Urumqi) at an angular
  resolution of 14$\farcm$5 ({\it right panels}) in the regions ``A''
  to ``E'' as marked in the {\it upper-left panel} of
  Fig.~\ref{fig:fg2}. To avoid confusion from ambient point-like
  sources, the sizes of region A and D are slightly reduced for the
  863-MHz and the 4.8-GHz images at 14$\farcm$5 resolution.}
\label{fig:fg3}
\end{figure*}

\subsection{Radio continuum spectrum}
The component-separation results can be regarded as a hint for the
emission nature of radio sources. The method is convenient for a quick
view of numerous radio structures in a vast sky area. However, we note
that the r.m.s noise level, i.e. 30 $-$ 50~mK\ $T_{b}$ of the
background emission in the separation results is higher than that in
the input survey map, i.e. $\sim$16~mK\ $T_{b}$ for the Effelsberg
1.4-GHz survey data. The decomposition method hence can work well on
structures with strong emission, but has some limitations for the
structural features that are too faint.

To confirm the overall spectrum nature of G213.0$-$0.6, we make
TT-plots \citep{Turtle62} between the observed Effelsberg 1.4-GHz and
the Urumqi 4.8-GHz data, and between the archived Effelsberg 863-MHz
\citep{Reich03} and the Urumqi 4.8-GHz data. The inclusion of the
Urumqi 4.8-GHz data, which was not available at the time of
\citet{Reich03}, significantly widens the frequency range than before
(863-MHz v.s. 2.7-GHz) and will lead to smaller uncertainties in
determining spectral indices.  Furthermore, the NVSS sources are first
overlaid onto the image, and five areas free from strong point-like
sources are selected (as marked with ``A'' to ``E'' in the {\it
  upper-left panel} of Fig.~\ref{fig:fg2}) for this purpose. We show
the TT-plot results for the five regions in Fig.~\ref{fig:fg3}. The
spectral indices of $\beta \sim -1.84$ to $-$2.17 ($\alpha = \beta +
2$) confirm and strongly support the flat-spectrum thermal emission
nature of G213.0$-$0.6. TT-plots toward the 30$\arcmin$ wide stripe
centered at $b = -10\arcmin$ as indicated in \citet{Reich03} is also
tested, but show scattered data points. We believe this is affected by
the existence of un-resolved point-like sources in the upper part of
G213.0$-$0.6 (see {\it upper-left panel} in Fig.~\ref{fig:fg2}). They
are very difficult to be removed in view of the 14$\farcm$5 resolution
of the 863-MHz data.
\begin{figure}
\centering
\includegraphics[angle=-90,width=0.46\textwidth]{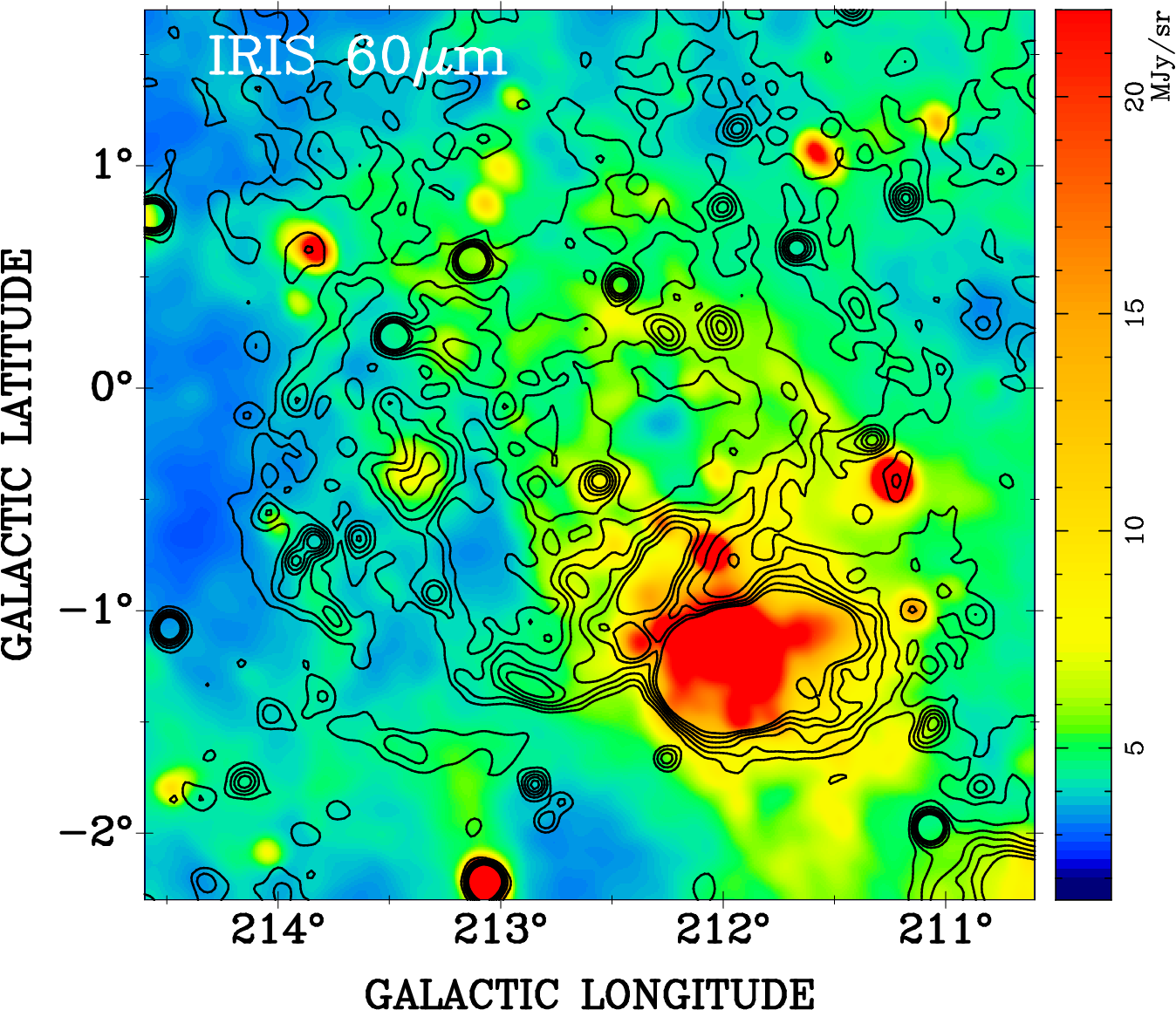}
\caption{IRIS 60$\mu$m infrared emission of the G213.0$-$0.6 area. The
  contours are from the Effelsberg 2.7-GHz radio continuum data
  smoothed to a resolution of 6$\arcmin$ to increase the
  signal-to-noise ratio.}
\label{fig:fg4}
\end{figure}

\section{Discussion}
\label{sect:dis}
\subsection{Morphology}
The morphology of G213.0$-$0.6 has been discussed in detail by
\citet{Stupar12}. The filamentary structures seen in both optical
H$\alpha$ and at radio frequencies are well correlated. However,
\ion{H}{II} regions and SNRs could share such a feature, e.g. the
\ion{H}{II} region Carina Nebula \citep{Smith10, Rebolledo21} and the
SNR S147 \citep{Greimel21, Xiao08}. The H$\alpha$ emission of
\ion{H}{II} regions comes from recombination of electrons and protons,
and the radio emission is from the free-free emission in the same
ionized gas. The H$\alpha$ and radio emission of SNRs originates from
the shock-excited interstellar medium and synchrotron emission from
shock-accelerated relativistic electrons, respectively. Therefore, a
solid classification cannot be made by morphological features, but
needs proofs, e.g. in optical emission-line ratios and radio continuum
spectrum. \citet{Stupar12} examined the infrared emission which traces
dust in the area of G213.0$-$0.6. They presented a color-coded (RGB,
red for radio continuum, green for 60$\mu$m infrared, and blue for
H$\alpha$) image and stated that there is no clear morphological
association between the infrared and the well-correlated optical-radio
emission. We present the IRIS 60$\mu$m image \citep{Miville05} in
Fig.~\ref{fig:fg4} and overlay the radio continuum contours of the
Effelsberg 2.7-GHz data \citep{Fuerst90}. Unlike the infrared emission
filling the entire \ion{H}{II} region SH 2-284, the 60$\mu$m emission
in the area of G213.0$-$0.6 is relatively weaker. However, it is
difficult to conclude that there is no morphological association,
because some infrared emission seems to be confined along the radio
boundary. It is also not clear that in case the dust was cleared away
by the stellar winds as G213.0$-$0.6 evolves, resulting in low
infrared emission in the field.

The X-ray source 1RXS J065049.7$-$003220 was proposed to be possibly
associated with G213.0$-$0.6 \citep{Stupar12}. Based on machine
learning, \citet{McGlynn04} developed the tool of
``ClassX''\footnote{https://heasarc.gsfc.nasa.gov/classx/} to classify
ROSAT sources \citep{Voges99}. The X-ray emission from 1RXS
J065049.7$-$003220 most probably stems from a
star. \citet{Haakonsen09} cross-matched the sources in the ROSAT All
Sky Survey Bright Source Catalog and the near-infrared sources from
the Two Micron All Sky Survey Point Source Catalog
\citep{Cutri03}. They found that 1RXS J065049.7$-$00322 may be
associated with a star in binary system. Hence, the X-ray emission
seen inside G213.0$-$0.6 seems not related.

\subsection{Distance and possible ionizing source}

The radial velocity (V$_{LSR}$) measured for the H$\alpha$ line in the
selected positions exhibits positive values averaged to around
20~km/s, except for the position G8 with approximately 6~km/s. The
origin of this singular V$_{LSR}$ value is ambiguous; it could reflect
the inherent filament motion in this segment of G213.0$-$0.6 or just
simply an overlap in a nearer distance, considering that it is located
far away from the major part of G213.6$-$0.6. According to
\citet{Reid14}, V$_{LSR}$ $\sim$ 20~km/s approximates the kinematic
distance of 1.9$^{+0.8}_{-0.7}$~kpc, placing G213.0$-$0.6 in the
Perseus Arm \citep{Hou21}, although with large uncertainties. The
physical size of G213.0$-$0.6 is therefore about 87~pc $\times$ 76~pc.

With the Gaia Early Data Release 3 \citep[EDR3,][] {Gaia22},
\citet{Xu21} used OB stars with parallax angles to outline the local
spiral structures within approximately 5~kpc from the Sun. We extract
OB stars in the field of G213.0$-$0.6 and the adjacent \ion{H}{II}
region SH 2-284 from \citet{Xu21}. About 30 stars are pinpointed, and
their distances are estimated using the parallax angles. There are
several OB-type stars in the area with distances ranging from 0.7 $-$
5~kpc (see Table~\ref{Tab:taba1}). Among them, two B-type stars, Gaia
DR3 3113244500922158464 and Gaia DR3 3113430696344419072 are located
in close proximity to the center of G213.0$-$0.6 (see the {\it
  upper-right panel} of Fig.~\ref{fig:fg2} and
Table~\ref{Tab:taba1}). Distinct distances can be directly estimated
from the parallax angle for the two stars: one at $\sim$0.7~kpc and
the other at $\sim$4.2~kpc. Neither of the two stars seems to be
related in view of distances. However, the distance of the latter,
Gaia DR3 3113430696344419072 reduces to $\sim$2.7~kpc after the
correction by \citet{Gaia22}, which largely narrows the gap and makes
it comparable to the kinematic distance inferred from the V$_{LSR}$ of
the H$\alpha$ filaments.

Within the core and outer regions of the \ion{H}{II} region SH 2-284,
a cluster of stars is found. The majority of these stars are
positioned at distances ranging from approximately 4 to 5~kpc. Some of
them are responsible for the ionization of SH 2-284
\citep[e.g][]{Negueruela15}. A B1II-type star is situated at the
center of the shell region between G213.0$-$0.6 and SH 2-284. The
distance for this star estimated from the parallax-angle measurement
is consistent with those in SH 2-284. Therefore, it might be possible
that the radio emission from the shell region originates from the
contribution of G213.0$-$0.6 itself and partially from the ionized gas
created by this star, assuming that it is indeed capable to generate
an \ion{H}{II} region from the same line of sight.
\section{Summary}
\label{sect:sum}
The low surface-brightness extended radio source G213.0$-$0.6 was
previously suggested to be an SNR. We revisit the nature of
G213.0$-$0.6 by combining the optical spectral line data together with
the radio continuum data. The new LAMOST observations of H$\alpha$,
[\ion{S}{II}], and [\ion{N}{II}] show lower spectral line ratios in
comparison to previous results, shaking the identify of G213.0$-$0.6
as being an SNR. Multi-wavelength radio continuum images are used to
separate non-thermal synchrotron emission from the thermal free-free
emission. Both the decomposition result and the additional
verification made by TT-plots confirm that G213.0$-$0.6 has a flat
spectrum, the key feature of thermal emission. OB-type stars are the
ionizing sources of \ion{H}{II} regions. The B-type star located at
$\ell = 212\fdg93$, $b = -0\fdg48$ might be responsible for the ionization 
of G213.0$-$0.6 due to the comparable distances inferred from the 
parallax angle (2.7~kpc) and from the $V_{LSR}$ measured for the
H$\alpha$ lines (1.9$^{+0.8}_{-0.7}$~kpc).

Combining all these evidence, we argue that G213.0$-$0.6 is possibly
not an SNR at a distance of 1~kpc, but an \ion{H}{II} region located
in the Perseus Arm. A further exploration of the optical lines of
[\ion{O}{I}], [\ion{O}{II}], [\ion{O}{III}], and H$\beta$ would help to
reach a solid confirmation about the genuine nature of G213.0$-$0.6.

\section*{Acknowledgements}
We thank the anonymous referee for helpful comments that improve the
paper.  XYG thanks the financial support by the National Natural
Science Foundation of China (No. 11833009), the National SKA program
of China No. 2022SKA0120103, the National Key R\&D Program of China
(No. 2021YFA1600401 and 2021YFA1600400), and the National Natural
Science Foundation of China (No. 11988101). CJW thanks the support by
the National Natural Science Foundation of China (Nos. 12090041,
12090040, 12073051).
This work presents results from the European Space Agency (ESA) space
mission Gaia. Gaia data are being processed by the Gaia Data
Processing and Analysis Consortium (DPAC). Funding for the DPAC is
provided by national institutions, in particular the institutions
participating in the Gaia MultiLateral Agreement (MLA).

\section*{Data Availability}
The data underlying this article will be shared on reasonable request
to the corresponding author.



\bibliographystyle{mnras}
\bibliography{example} 



\appendix


\section{Auxiliary Table}

\begin{table*}
\begin{center}
  \caption{OB stars around the center of G213.0$-$0.6, the southwest
    shell and the \ion{H}{II} region SH 2-284.  The sequential number
    and the names of the stars are listed in the 1st and 2nd
    column. The Galactic coordinates are shown in the 3rd and 4th
    column. The parallax angle from \citet{Gaia22}, the distance
    derived directly from the parallax angle (d = 1 A.U. /
    tan(parallax angle)), and the distance after correction by
    \citet{Gaia22} are presented in 5th $-$ 7th column. The spectral
    type of the stars extracted from \citet{Xu21} are given in the 8th
    column.}
    \label{Tab:taba1}
\begin{tabular}{clcccccc}
\hline\hline\noalign{\smallskip}
\multicolumn{1}{c}{No.}  &\multicolumn{1}{c}{Name}  & \multicolumn{1}{c}{Galactic longitude} & \multicolumn{1}{c}{Galactic latitude}   & \multicolumn{1}{c}{Plx} & \multicolumn{1}{c}{Distance} & \multicolumn{1}{c}{Distance$^{*}$} & \multicolumn{1}{c}{SpecType} \\
&  & \multicolumn{1}{c}{($\degr$)}  & \multicolumn{1}{c}{($\degr$)}  & \multicolumn{1}{c}{(mas)} & \multicolumn{1}{c}{(pc)} & \multicolumn{1}{c}{(pc)} & \\
\hline\noalign{\smallskip}
   & {\bf G213.0$-$0.6}                &            &          &       &       &            &\\
1  & Gaia DR3 3113244500922158464      & 212.93     & $-$0.48  &0.2362 & 4235  &  2768      & B0: \\
2  & Gaia DR3 3113430696344419072      & 212.90     & $-$0.53  &1.4960 &  669  &  $\cdots$  & B2V \\
\hline
   & {\bf Southwest shell region}                &            &          &       &       &           & \\
3  &Gaia DR3 3113482545188572928      & 212.33     & $-$0.91  &0.2221 & 4504  &  3894     & B1II \\
\hline
   &{\bf SH 2-284 and surroundings}   &            &          &       &       &           & \\
4  &Gaia DR3 3125505326881596544      & 212.04     & $-$1.28  &0.2355 & 4248  &  3416     & B2V \\
5  &Gaia DR3 3125505563098855552      & 212.01     & $-$1.32  &0.2131 & 4694  &  3942     & B2III \\
6  &Gaia DR3 3125505670478616192      & 212.00     & $-$1.31  &0.2281 & 4386  &  3303     & O7Vz \\
7  &Gaia DR3 3125513878155537024      & 211.96     & $-$1.15  &0.1818 & 5503  &  $\cdots$ & B0Ve \\
8  &Gaia DR3 3119511442321221120      & 211.95     & $-$1.63  &0.1914 & 5227  &  2819     & B2IV \\
9  &Gaia DR3 3119511854638059136      & 211.94     & $-$1.60  &0.2389 & 4187  &  4536     & B2IV \\
10 &Gaia DR3 3125518039984467328      & 211.93     & $-$1.39  &0.2479 & 4035  &  2943     & O9.7V + B \\
11 &Gaia DR3 3125525629188072704      & 211.86     & $-$1.31  &0.2010 & 4977  &  3360     & B2IV \\
12 &Gaia DR3 3125522742970050304      & 211.81     & $-$1.43  &0.1579 & 6336  &  4336     & B1Ve \\
13 &Gaia DR3 3125522197512839680      & 211.77     & $-$1.50  &0.2663 & 3757  &  3637     & B1.5V \\
14 &Gaia DR3 3125522197512841856      & 211.77     & $-$1.51  &0.2253 & 4440  &  5223     & B1.5V \\
15 &Gaia DR3 3125557828561213568      & 211.72     & $-$1.00  &0.2237 & 4472  &  3640     & B2III \\
16 &Gaia DR3 3125569545232312320      & 211.70     & $-$1.53  &0.2220 & 4506  &  3309     & B2IV \\
17 &Gaia DR3 3125603973689913856      & 211.63     & $-$1.17  &0.2201 & 4545  &  4888     & O8.5Ib(f) \\
18 &Gaia DR3 3125580780866556416      & 211.63     & $-$1.25  &0.5957 & 1679  &  1357     & B0.5IV \\
19 &Gaia DR3 3125585969186497280      & 211.48     & $-$1.26  &0.2400 & 4168  &  3756     & B1.5II-III \\
20 &Gaia DR3 3125597681559414784      & 211.46     & $-$1.28  &0.2481 & 4032  &  2807     & O8Vz + B0:V: \\
\noalign{\smallskip}\hline
\end{tabular}
\end{center}

\end{table*}

\bsp	
\label{lastpage}
\end{document}